\documentclass{rmf-d}
\usepackage{nopageno,rmfbib,multicol,wrapfig,times,epsf,amsmath,amssymb,cite,here}
\usepackage[latin1]{inputenc}
\usepackage[]{caption2}
\usepackage{graphicx}
\def\rmfcornisa{Research OR Education of Physics \hfill\rmf\ {\bf ??} (*?*) ???--???
\hfill MES? A\~NO?}

\def\rmfcintilla{{\it Rev.\ Mex.\ Fis.\/} {\bf ??} (*?*) (????) ???--???}
\clearpage \rmfcaptionstyle 
\begin{document}
\title{Large-x power laws of parton distributions remain inconclusive
\vspace{-6pt}}
\author{Aurore~Courtoy }
\address{Instituto de F\'isica, Universidad Nacional Aut\'onoma de M\'exico\\
Apartado Postal 20-364, 01000 Ciudad de M\'exico, Mexico }
\author{ Pavel M. Nadolsky       }
\address{  Department of Physics, Southern Methodist University, Dallas, TX 75275-0181, U.S.A.   }

\maketitle
\recibido{day month year}{day month year
\vspace{-12pt}}
\begin{abstract}
\vspace{1em} Focusing on hadron scattering at large partonic momentum fractions $x$, we compare nonperturbative QCD predictions for the asymptotic behavior of DIS structure functions and parton distribution functions (PDFs) to the $x$ and $Q$ dependence of phenomenological PDFs. In the CT18 NNLO global QCD analysis, higher-order radiative contributions and functional mimicry of PDF parametrizations result in about one unit of uncertainty in the effective powers of $(1-x)$ falloff of nucleon PDFs. Similar uncertainties are present in the case of the pion PDF, an object of growing interest in phenomenology. \vspace{1em}
\end{abstract}
\keys{parton distribution functions, hadron structure, quark counting rules   \vspace{-4pt}}
\vspace{1\baselineskip}

\begin{multicols}{2}
\section{Quark counting rules in the large-$x$ limit}

Parton Distribution Functions (PDFs) are the key inputs to unveiling the hadron structure in high-energy scattering. At sufficiently large partonic momentum fractions, typically $x\gtrsim 0.1$, these nonperturbative QCD functions can be evaluated in theoretical models, effective theories, in lattice QCD or in the continuum, or alternatively determined by global analyses from experimental observations. While the factorization scale dependence of the PDFs is dictated by DGLAP evolution equations (starting from an initial scale of order 1 GeV), their dependence on the fraction of longitudinal momentum $x$ of the parent hadron is not predicted by perturbative QCD. Together with the first QCD principles, a few rules guide the shape of the PDFs in some specific regions: positivity constraints and quark counting rules (QCRs) are such examples.

In these proceedings, we summarize a recent study~\cite{Courtoy:2020fex} of phenomenological implications of the QCRs in a broad range of scattering processes accessible in the global QCD analysis of PDFs. The QCRs have been first derived for DIS structure functions at large $x$ in early QCD models~\cite{Ezawa:1974wm,Farrar:1975yb,Berger:1979du,Soper:1976jc}.
The QCRs reflect kinematic properties of the lowest-order scattering amplitudes for DIS cross sections in the regime dominated by semi-hard gluon exchanges. In this regime, the DIS structure functions decrease with Bjorken $x_B$ as $(1-x_B)^{2 n_s-1+2|\lambda_q-\lambda_A|}$,  where $n_s$ is the number of quark spectators, and $\lambda_{q,A}$ are the helicity of the active quark and parent hadron, respectively. 

When the QCRs are extended from the large-$x_B$ structure functions to the large-$x$ {\it universal} parton distribution functions, various aspects of QCD must be taken into account. Those include QCD factorization, radiative contributions at the NLO and beyond, and, at the low scale values that are most relevant to accessing the large-$x$ regime, target-mass corrections as well as all-order resummation. Also, the QCRs are motivated by the kinematics of the lowest-order semi-hard QCD interactions in the low to mid-$Q$ regimes, supplemented by the DGLAP evolution of the PDFs to approximate the higher-order radiative contributions. In the comparisons to the nonperturbative predictions, on the other hand, one aims to bridge these {\it pheno} PDFs, expressed in terms of the perturbative degrees of freedom, to distribution functions evaluated in the low-energy approaches operating with nonperturbative representations of QCD. The task is gargantuan, yet future experiments and theoretical efforts should lead us in that direction. 

\section{Polynomial mimicry}
When finding the PDFs ${\cal F}(x, Q^2)$ from a phenomenological analysis, many assumptions are usually made. [Our notations omit the PDF flavor indices for brevity.] In particular, the $x$ dependence at the starting evolution scale $Q_0$ is parameterized by a functional form with fitted free parameters. A typical parametrization of a proton PDF modulates a baseline function that drives the behavior at $x\to 0$ and $x\to 1$ by a smooth function $\Phi$, as
\begin{equation}
{\cal F}(x, Q^2_0)=x^{A_{1}}(1-x)^{A_{2}}\times \Phi(x; A_3,...,\ A_n).
\label{eq:ff}
\end{equation}
Similar parametrizations can and have been adopted for analyses of the pion PDF, see {\it e.g.}~\cite{Barry:2018ort,Novikov:2020snp,Barry:2021osv}. The best-fit values of the free parameters $A_{1,2}$ can be examined to learn about the PDF dynamics in the asymptotic limits~\cite{Ball:2016spl,Courtoy:2020fex}. In particular, the parameter $A_2$ from a fit is often interpreted in literature as the primordial exponent of the large-$x$ falloff, but this can be misleading. 
It is important to note here that discrete data points in a finite $x$ range are compatible with more than one continuous functional form: that is, it is easy to show mathematically \cite{Courtoy:2020fex} that infinitely many functions $\Phi(x; A_3,...,\ A_n)$ result in the same quality of a fit to the data at hand. The best-fit parameters $A_2$ determined this way are correlated with all parameters $A_n$ in $\Phi(x; A_3,...,\ A_n)$. Therefore, it cannot be proved that experimental data demand a $(1-x)^{A_2}$ fall-off, where $A_2$ would be larger than $3$ for a proton and $2$ for a pion. We call this feature a {\it polynomial mimicry}~\cite{Courtoy:2020fex}. It can be demonstrated with exact interpolations based on B\'ezier curves, each rendering a unique polynomial solution for the chosen set of $(n+1)$ sampled points.

\section{The case of the pion PDFs \label{sec:pion}}

Can a set of discrete data points that sample a (pion) PDF demand the minimal suppression power for this (pion) PDF? One possible strategy to address this question is to examine the monomial expansions of the B\'ezier curves that fit the data \cite{Courtoy:2020fex}, 
\begin{eqnarray}
 {\cal B}^{(n)}(x) = \sum_{l=0}^n \bar{c}_l\  (1-x)^l.
  \label{eq:Bnonemx}
\end{eqnarray}
We can raise this question not only for a pheno PDF found from experimental measurements, but also when the PDF is predicted by a theoretical calculation, as long as it is presented in the form given by Eq.~(\ref{eq:ff}). 

The monomial expansion shows that the parameters $\bar c_l$ of the reconstructed functional forms depend on the $x$ range spanned by the data. More strikingly, we have observed a mixing of expansion coefficients, $\bar{c}_l$ with various $l$, as well as the appearance of spurious coefficients. Mimicry therefore dilutes the  connection between the data and theoretical truth. In other words, the available data are compatible, within a given uncertainty/confidence interval, with multiple polynomial solutions, including those suggested by the early QCD and nonperturbative models. This eliminates the concern about a possible contradiction between pion data and theory raised in Ref.~\cite{Cui:2021mom}. 

\begin{figure}[H]
\includegraphics[width=1\linewidth]{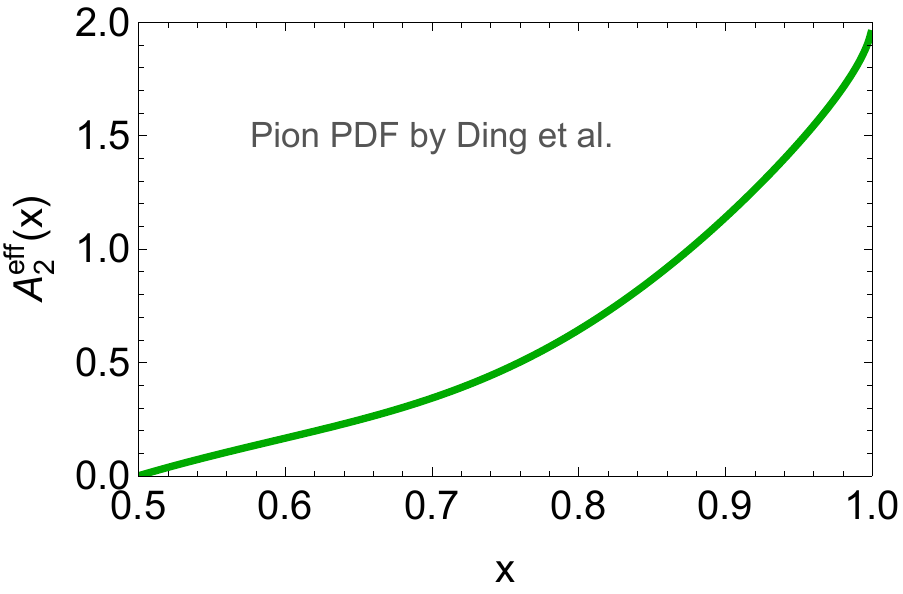}
\caption{Effective $(1-x)$ exponent $A_2^{\mbox{eff}}(x)$ for the pion PDF of Ref.~\cite{Ding:2019lwe}.}
\label{fig:EffA2Ding}
\end{figure}

For the pion asymptotics to be more meaningfully interpreted at large $x$, data at $x\gtrsim 0.9$ are necessary to minimize that mixing of the B\'ezier coefficients. Until then, only increased uncertainties at large $x$ can appropriately represent the multiple choices for the functional forms in the region $x\to 1$. This is best understood by considering the {\it effective exponent} at large-$x$ values, defined as
\begin{equation}
A_2^{\rm eff}\left[{\cal F}(x,Q^2)\right]\equiv \frac{\partial \ln\left({\cal F}(x, Q^2)\right)}{\partial \ln\left(1-x\right)}=A_2\left[{\cal F}\right] +\mbox{correction}
\;,
\label{eq:A2eff}
\end{equation}
 i.e., the logarithmic derivative of the PDF ${\cal F}(x, Q^2)$. In Fig.~\ref{fig:EffA2Ding}, we show $A_2^{\rm eff}$ for the pion PDF obtained through a Dyson-Schwinger formalism~\cite{Ding:2019lwe}, given at a low hadronic scale of a few hundred MeV. Here the input PDF is of the form $(1-x)^2 \Phi^\prime(x)$. 
 According to the figure, $A_2^{\rm eff}$ approaches the expected value of 2 in the limit $x\to 1$ and quickly drops below 2 at smaller $x$ values. In the phenomenological analyses, however, the $x$ values above 0.9 are the most challenging, although not entirely hopeless in the pion case. These values, as already stated above, will determine the relevant corrections that must be accounted for. 
 
 A thorough study of large-$x$ resummation for the pion PDFs~\cite{Barry:2021osv} shows that the effective exponent is compatible with the QCR expectation {\it within uncertainties}. Those uncertainties are not small and reflect the choice of resummation technique. It is then a theoretical error like the choice of the parametrization form, which has not yet been studied for the pion.
 
 Another source of theoretical uncertainties comes when considering  predictions at the hadronic scale. At very low scales, owing to the dynamics of QCD, various nonperturbative manifestations might come into play, such as the broadening of the pion PDF due to the breaking of chiral symmetry. The latter, in turn, cannot be univocally distinguished from the QCR behavior based on a single functional form. QCD evolution to mid-energies, where QCR should be most relevant~\cite{Courtoy:2020fex,Cui:2021mom}, softens the shape that had been acquired at a low energy through mass-generation effects.

 Similar conclusions regarding the determination of the large-$x$ exponent of the pion PDF have been reached by lattice practitioners~\cite{Gao:2020ito,Cichy:2021lih}, altogether indicating that the determination of the primordial large-$x$ falloff of PDF is an {\it ill-posed problem}. Nevertheless, the effective exponent defined in Eq.~(\ref{eq:A2eff}) can be a helpful metric for the comparisons with theory, even if it needs not coincide with the QCR prediction at accessible $x$. 

\section{The case of the proton PDFs \label{sec:proton}}

We will now illustrate the behavior of the effective asymptotic exponents for proton scattering, for which both precise theoretical and experimental inputs already exist. 

\begin{figure}[H]
\includegraphics[width=1\linewidth]{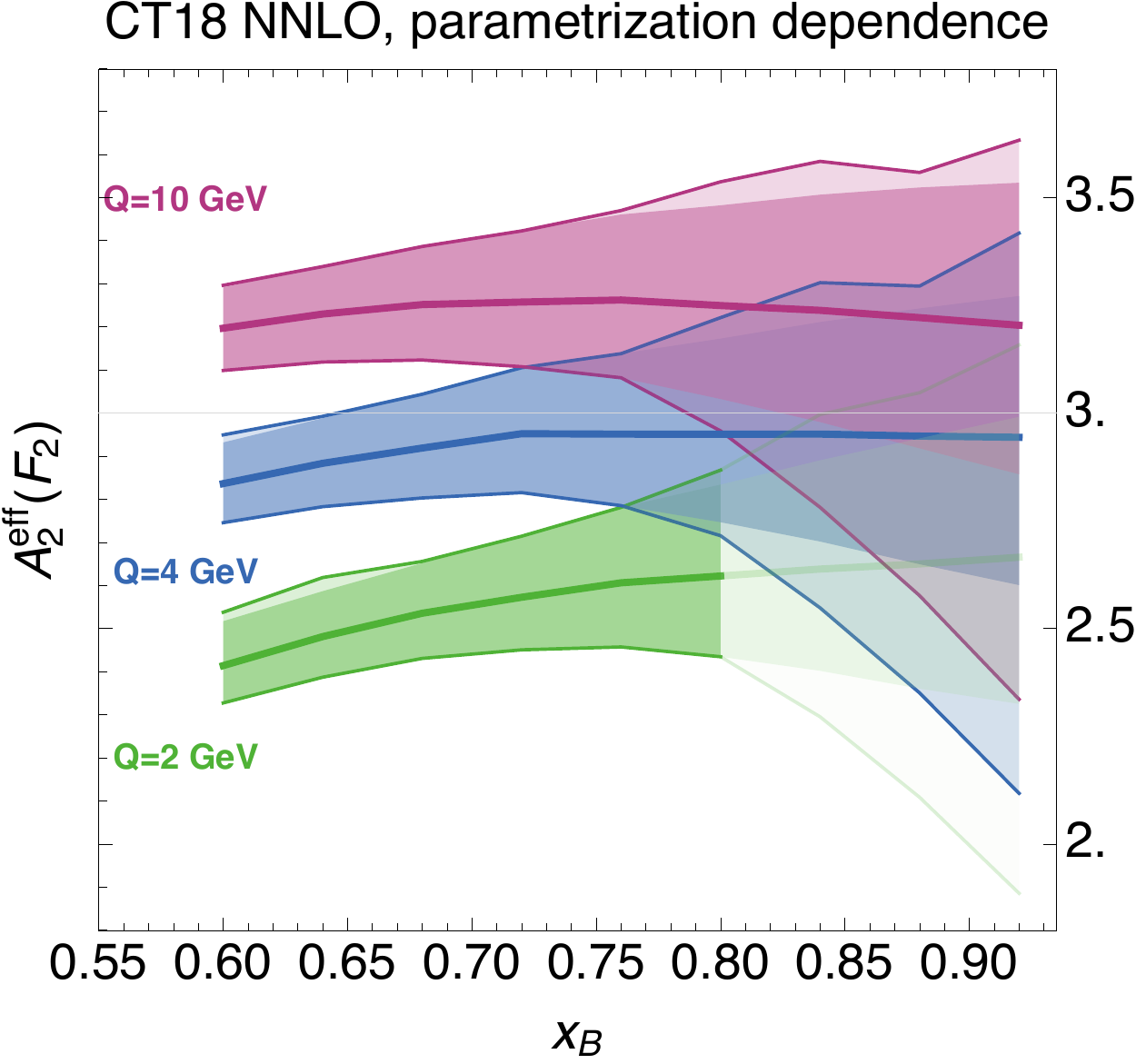}
\caption{Effective $(1-x)$-exponent $A_2^{\mbox{eff}}(x)$ for the proton structure function $F_2(x,Q^2)$ based on the PDFs of Ref.~\cite{Hou:2019efy}, as a function of $x_{\rm B}$ and for the $Q$ values of $2,\, 4$ and $10$ GeV represented in green, blue and magenta, respectively. The central curve of each error band represents the CT18 NNLO central value, the dark shaded band is the asymmetric Hessian error \cite{Nadolsky:2001yg,Lai:2010vv} at the 68\% probability level. The extreme curves correspond to the envelope of the Hessian and parametrization uncertainties estimated based on 363 functional forms. The transparent part of the $Q=2$ GeV band corresponds to the region with $W^2<2m_p^2$, approximately corresponding to the resonance region in DIS. The reference prediction from the QCRs is shown by a line at $A_2^{\mbox{\tiny eff}}(F_2)=3$.}
\label{fig:rainbow}
\end{figure}

We will use the CT18 ensemble of NNLO proton PDFs \cite{Hou:2019efy}. The CT18NNLO ensemble has been extended to include $363$ functional forms as part of the study of its uncertainty bands. In Ref.~\cite{Courtoy:2020fex}, we have used those alternative functional forms to study the behavior of the falloff of the structure functions and PDFs at large $x$. The DIS structure function $F_2(x,Q^2)$ was found to be compatible with the quark counting rule expectation of $A_2^{\rm eff}(F_2)=3$ as $x\to 1$, within error bands, as shown in Fig.~\ref{fig:rainbow}. The determination of PDF uncertainty is a multifactorial process~\cite{Kovarik:2019xvh}: one of these factors is the choice of functional form, which is represented by the light-shaded areas in the aforementioned plot. In particular, the error bands must be large in the regions with scarce data, such as $x\to 1$. 

In the same reference \cite{Courtoy:2020fex}, we have shown that the QCR predictions are not fulfilled for the sea and the gluon PDFs at low scales around the CT18's value $Q_0=1.3$ GeV. However,
the effective exponents of the PDFs evolve with $Q^2$ in accordance with DGLAP evolution, which is known very accurately to NNLO in $\alpha_s$.  The correlated behavior of the valence or gluon and antiquark PDFs becomes manifest when $Q$ increases.

\begin{figure}[H]
\includegraphics[width=\linewidth]{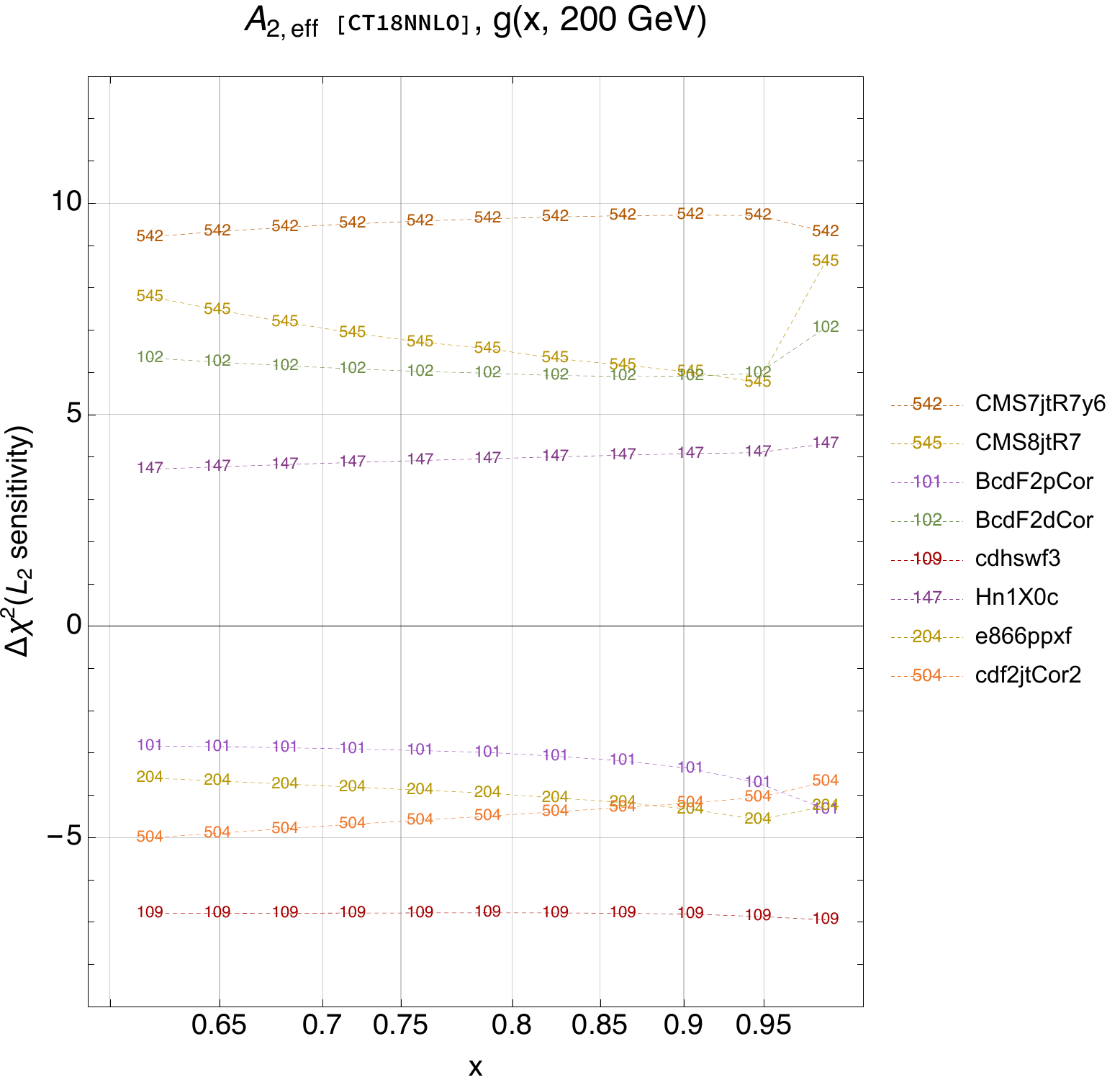}
\caption{$L_2$-sensitivity plot for the gluon CT18NNLO PDF at 200 GeV.}
\label{fig:L2gluon}
\end{figure}

Since the rate of $Q$ evolution depends weakly on the parametrization form, knowledge of QCD evolution opens interesting opportunities for testing $A_2^{\rm eff}$ at nonperturbative $Q\sim 1$ GeV by using large-$x$ constraints at very high $Q$ from collider experiments, such as the ZEUS measurements \cite{ZEUS:2020ddd} of DIS at $Q\approx 100$ GeV~\cite{Courtoy:2021xpb}. The impact of high-energy data can be studied through the $L_2$ sensitivity analysis: $S_{2,L}(E)$ for experiment $E$ is the estimated $\Delta \chi^2$ for this experiment when a quantity $F(x, Q)$ increases by the $+68\%$ c.l. Hessian PDF uncertainty~\cite{Wang:2018heo,Hobbs:2019gob}.
In Fig.~\ref{fig:L2gluon}, we show $S_{2,L}(E)$ corresponding to the effective exponent $A_2^{\rm eff}$ for the gluon at 200 GeV: at large $x$, the effective exponent for the gluon falloff is presently determined by a trade-off between jet, neutrino-nucleus DIS, DIS and DY data. We see some opposing pulls on $A_2^{\rm eff}$ for the gluon between these groups of experiments, suggesting that yet uncontrolled effects may lead to a mild inconsistency. 

Another phenomenologically relevant combination for testing the nonperturbative dynamics is the $d/u$ ratio, discussed, {\it e.g.}, in Ref.~\cite{Accardi:2021ysh}. 
As both the up and down valence PDFs evolve according to the same non-singlet DGLAP evolution, the $Q^2$ dependence of their effective exponents cancels in a wide $x$ range~\cite{Courtoy:2021xpb}. The CT18 parametrizations assume the ratio $A_{2,u_{\rm V}}/A_{2,d_{\rm V}}$ of the fitted exponents in Eq.~(\ref{eq:ff}) to be equal to $1$ to ensure that the $d/u$ remains finite. By studying this ratio for 363 trial parametrizations of CT18 NNLO PDFs versus $Q$, we observe that the effective exponents at $x < 1$  deviate from the fitted ones. The ratio is nearly invariant under DGLAP evolution, so it can be determined at ZEUS or the LHC and provide accurate insights about the flavor composition in the valence sector at $Q\sim 1$ GeV. 

\section{Conclusions}

In these proceedings, we have explored the phenomenology of PDFs at large $x$ in view of the constraints given by the quark counting rules. We have argued that the effective exponents predicted by the theoretical methods for valence quark PDFs are consistent with experimental observations within one-two units, namely: $0\lesssim A_2^{\rm eff, \pi}\lesssim 4$ and $1\lesssim A_2^{\rm eff, P}\lesssim 5$. Tests to a higher accuracy have been challenging because of two classes of uncertainties:

\begin{enumerate}
\item[1.] Higher-order and higher-power QCD contributions are often not assessed. To know the associated uncertainties, perturbatively stable factorization with universal PDFs must be demonstrated in the relevant kinematic regions for the examined observables.
\end{enumerate}
\begin{enumerate}
\item[2.] Determination of the asymptotic exponents relies on end-point extrapolations, analogously to measurements of a neutrino mass in weak decays or lattice extrapolations to the physical pion mass. In such determinations, the estimated derivatives are highly correlated with the end-point value of the extrapolated function as well as its higher-order derivatives. Theoretical or phenomenological estimates of $A_2$ involving inter/extrapolation are sensitive to user-chosen high-power terms of the extrapolating polynomial.  Mathematical underpinnnings of this functional mimicry are addressed in \cite{Courtoy:2020fex}.
\end{enumerate}
The published analyses most often neglect these systematic uncertainties, which are of theoretical rather than experimental nature. Control of these uncertainties is critical for testing the power laws incisively. Sections~3 and 4 elaborate on these issues in the contexts of the pion and proton analyses.

\section*{Acknowledgements}
AC is supported by UNAM Grant No. DGAPA-PAPIIT IA101720 and CONACyT Ciencia de Frontera 2019 No.~51244 (FORDECYT-PRONACES). PN is partially supported by the U.S.~Department of Energy under Grant No.~DE-SC0010129.

\end{multicols}
\medline
\begin{multicols}{2}

%\bibliographystyle{unsrt}
%\bibliography{biblio.bib}

\begin{thebibliography}{99}
%\cite{Courtoy:2020fex}
\bibitem{Courtoy:2020fex}
A.~Courtoy and P.~M.~Nadolsky,
%``Testing momentum dependence of the nonperturbative hadron structure in a global QCD analysis,''
Phys. Rev. D \textbf{103}, 054029 (2021)
doi:10.1103/PhysRevD.103.054029
[arXiv:2011.10078 [hep-ph]].
%8 citations counted in INSPIRE as of 20 Dec 2021

%\cite{Ezawa:1974wm}
\bibitem{Ezawa:1974wm}
Z.~F.~Ezawa,
%``Wide-Angle Scattering in Softened Field Theory,''
Nuovo Cim. A \textbf{23}, 271 (1974),\\
doi:10.1007/BF02739483
%115 citations counted in INSPIRE as of 20 Dec 2021

%\cite{Farrar:1975yb}
\bibitem{Farrar:1975yb}
G.~R.~Farrar and D.~R.~Jackson,
%``Pion and Nucleon Structure Functions Near x=1,''
Phys. Rev. Lett. \textbf{35}, 1416 (1975)
doi:10.1103/PhysRevLett.35.1416
%515 citations counted in INSPIRE as of 20 Dec 2021

%\cite{Berger:1979du}
\bibitem{Berger:1979du}
E.~L.~Berger and S.~J.~Brodsky,
%``Quark Structure Functions of Mesons and the Drell-Yan Process,''
Phys. Rev. Lett. \textbf{42}, 940 (1979)
doi:10.1103/PhysRevLett.42.940
%355 citations counted in INSPIRE as of 20 Dec 2021

%\cite{Soper:1976jc}
\bibitem{Soper:1976jc}
D.~E.~Soper,
%``The Parton Model and the Bethe-Salpeter Wave Function,''
Phys. Rev. D \textbf{15}, 1141 (1977),\\
doi:10.1103/PhysRevD.15.1141
%245 citations counted in INSPIRE as of 22 Dec 2021

%\cite{Ball:2016spl}
\bibitem{Ball:2016spl}
R.~D.~Ball, E.~R.~Nocera and J.~Rojo,
%``The asymptotic behaviour of parton distributions at small and large $x$,''
Eur. Phys. J. C \textbf{76}, 383 (2016)
doi:10.1140/epjc/s10052-016-4240-4
[arXiv:1604.00024 [hep-ph]].
%56 citations counted in INSPIRE as of 21 Dec 2021

%\cite{Barry:2018ort}
\bibitem{Barry:2018ort}
P.~C.~Barry, N.~Sato, W.~Melnitchouk, C.-R.~Ji,
%``First Monte Carlo Global QCD Analysis of Pion Parton Distributions,''
Phys. Rev. Lett. \textbf{121}, 152001 (2018)
doi:10.1103/PhysRevLett.121.152001
[arXiv:1804.01965 [hep-ph]].
%95 citations counted in INSPIRE as of 13 Dec 2021

%\cite{Novikov:2020snp}
\bibitem{Novikov:2020snp}
I.~Novikov \textit{et al.}
%``Parton Distribution Functions of the Charged Pion Within The xFitter Framework,''
Phys. Rev. D \textbf{102}, 014040 (2020)
doi:10.1103/PhysRevD.102.014040
[arXiv:2002.02902 [hep-ph]].
%50 citations counted in INSPIRE as of 20 Dec 2021

%\cite{Barry:2021osv}
\bibitem{Barry:2021osv}
P.~C.~Barry \textit{et al.} [Jefferson Lab Angular Momentum (JAM)],
%``Global QCD Analysis of Pion Parton Distributions with Threshold Resummation,''
Phys. Rev. Lett. \textbf{127},  232001 (2021)
doi:10.1103/PhysRevLett.127.232001
[arXiv:2108.05822 [hep-ph]].
%6 citations counted in INSPIRE as of 20 Dec 2021

%\cite{Cui:2021mom}
\bibitem{Cui:2021mom}
Z.~F.~Cui \textit{et al.},
%``Concerning pion parton distributions,''
[arXiv:2112.09210 [hep-ph]].
%0 citations counted in INSPIRE as of 23 Dec 2021

%\cite{Ding:2019lwe}
\bibitem{Ding:2019lwe}
M.~Ding \textit{et al.},
%``Symmetry, symmetry breaking, and pion parton distributions,''
Phys. Rev. D \textbf{101}, 054014 (2020)
doi:10.1103/PhysRevD.101.054014
[arXiv:1905.05208 [nucl-th]].
%62 citations counted in INSPIRE as of 23 Dec 2021

%\cite{Gao:2020ito}
\bibitem{Gao:2020ito}
X.~Gao \textit{et al.},
%``Valence parton distribution of the pion from lattice QCD: Approaching the continuum limit,''
Phys. Rev. D \textbf{102}, 094513 (2020)
doi:10.1103/PhysRevD.102.094513
[arXiv:2007.06590 [hep-lat]].
%39 citations counted in INSPIRE as of 21 Dec 2021

%\cite{Cichy:2021lih}
\bibitem{Cichy:2021lih}
K.~Cichy,
%``Progress in $x$-dependent partonic distributions from lattice QCD,''
[arXiv:2110.07440 [hep-lat]].
%7 citations counted in INSPIRE as of 21 Dec 2021

%\cite{Hou:2019efy}
\bibitem{Hou:2019efy}
T.-J.~Hou \textit{et al.}
%``New CTEQ global analysis of quantum chromodynamics with high-precision data from the LHC,''
Phys. Rev. D \textbf{103}, 014013 (2021)
doi:10.1103/PhysRevD.103.014013
[arXiv:1912.10053 [hep-ph]].
%183 citations counted in INSPIRE as of 24 Dec 2021

%\cite{Kovarik:2019xvh}
\bibitem{Kovarik:2019xvh}
K.~Kova\v{r}\'\i{}k, P.~M.~Nadolsky and D.~E.~Soper,
%``Hadronic structure in high-energy collisions,''
Rev. Mod. Phys. \textbf{92}, 045003 (2020)
doi:10.1103/RevModPhys.92.045003
[arXiv:1905.06957 [hep-ph]].
%42 citations counted in INSPIRE as of 23 Dec 2021

%\cite{Nadolsky:2001yg}
\bibitem{Nadolsky:2001yg}
P.~M.~Nadolsky and Z.~Sullivan,
%``PDF Uncertainties in WH Production at Tevatron,''
eConf \textbf{C010630}, P510 (2001)
[arXiv:hep-ph/0110378 [hep-ph]].
%61 citations counted in INSPIRE as of 24 Dec 2021

%\cite{Lai:2010vv}
\bibitem{Lai:2010vv}
H.-L.~Lai \textit{et al.},
%``New parton distributions for collider physics,''
Phys. Rev. D \textbf{82}, 074024 (2010)
doi:10.1103/PhysRevD.82.074024
[arXiv:1007.2241 [hep-ph]].
%3352 citations counted in INSPIRE as of 22 Dec 2021

%\cite{ZEUS:2020ddd}
\bibitem{ZEUS:2020ddd}
I.~Abt \textit{et al.} [ZEUS],
%``Study of proton parton distribution functions at high $x$ using ZEUS data,''
Phys. Rev. D \textbf{101}, 112009 (2020)
doi:10.1103/PhysRevD.101.112009
[arXiv:2003.08742 [hep-ex]].
%7 citations counted in INSPIRE as of 03 Dec 2021

%\cite{Courtoy:2021xpb}
\bibitem{Courtoy:2021xpb}
A.~Courtoy and P.~M.~Nadolsky,
%``Nucleon and pion PDFs: large-x asymptotics meets functional mimicry,''
[arXiv:2108.04122 [hep-ph]].
%2 citations counted in INSPIRE as of 03 Dec 2021

%\cite{Wang:2018heo}
\bibitem{Wang:2018heo}
B.~T.~Wang \textit{et al.},
%``Mapping the sensitivity of hadronic experiments to nucleon structure,''
Phys. Rev. D \textbf{98}, 094030 (2018)
doi:10.1103/PhysRevD.98.094030
[arXiv:1803.02777 [hep-ph]].
%26 citations counted in INSPIRE as of 24 Dec 2021

%\cite{Hobbs:2019gob}
\bibitem{Hobbs:2019gob}
T.~J.~Hobbs \textit{et al.},
%``Charting the coming synergy between lattice QCD and high-energy phenomenology,''
Phys. Rev. D \textbf{100}, 094040 (2019)
doi:10.1103/PhysRevD.100.094040
[arXiv:1904.00022 [hep-ph]].
%18 citations counted in INSPIRE as of 03 Dec 2021

%\cite{Accardi:2021ysh}
\bibitem{Accardi:2021ysh}
A.~Accardi, T.~J.~Hobbs, X.~Jing and P.~M.~Nadolsky,
%``Deuterium scattering experiments in CTEQ global QCD analyses: a comparative investigation,''
Eur. Phys. J. C \textbf{81},  603 (2021)
doi:10.1140/epjc/s10052-021-09318-y
[arXiv:2102.01107 [hep-ph]].
%5 citations counted in INSPIRE as of 24 Dec 2021
\end{thebibliography}

\end{multicols}
\end{document}